\documentclass[a4paper,10pt]{article}
\usepackage[utf8x]{inputenc}

\title{Absence of Black Holes  Information Paradox in Group Field Cosmology}
\author{ Mir Faizal \\
Mathematical Institute, University of Oxford
\\ Oxford
OX1 3LB, United Kingdom \\mir@maths.ox.ac.uk
 }
\date{}
\begin{document}

\maketitle

\begin{abstract}
 In this paper we  will analyse the black hole information paradox  in group field cosmology. 
We will first construct a group field cosmology with third quantized gauge symmetry. 
Then we will argue that that in this group field cosmology the process that  change the topology of spacetime 
are unitarity process. Thus, the information paradox  from this perspective  appears only because 
 we are using a second quantized formalism to explain a third quantized process.
 A similar paradox would also occur if we analyse a second quantized process in 
  first quantized  formalism. 
Hence, we will demonstrated that in reality there is no information paradox 
but only a breakdown of the second quantized formalism.  
\end{abstract}

\section{Introduction}
Semi-classical gravity predicts that a black hole will  evaporate by radiating a thermal radiation called the 
 Hawking radiation \cite{1}-\cite{1ab}. Furthermore, as this  radiation is thermal it cannot represent 
the information inside the black hole. Thus, 
after the black hole have evaporated completely, its wave function 
seems to disappear. 
 \cite{2}-\cite{2ab}. This means that  after a certain point 
the wave function of the system is not able to describe the system.
This lead to a non-unitary evolution for the system. 
 Various attempts have been made
 to resolve this paradox \cite{2b}-\cite{2c}. 
The most famous  of them being the  $AdS/CFT$ correspondence \cite{f0}-\cite{f1}.
This correspondence states that the superconformal field theory living on the boundaries 
of anti-de Sitter spacetime is dual to the classical gravity in its  bulk. Now as 
 there is no information loss 
on the $CFT$, so by this duality, it is asserted that
there would be no information loss on the $AdS$ side as well. Thus, it is argured that the
 black hole evaporation is also unitarity \cite{f2}-\cite{f3}.  Even though this argument is 
the best argument constructed so far, there still is a debate about the information loss in black holes. 
This is because a mechanism for the information to escape out of the black hole is still not known. 
Various theories  have been proposed for explaining this. 
These include    leaking of information out a
 of black hole  \cite{sl}-\cite{sl1}, 
information getting stored in a Planck sized remnant \cite{plz}-\cite{plz1}, 
information being encoded in the correlations between future and past \cite{e1}, 
and information escaping to a baby universe that is separates from our own universe \cite{bu}-\cite{bu1}. 
But none of these proposals have provided a fully satisfying answer to 
the information paradox and so despite all these approaches  the paradox is still not fully resolved. 
In this letter we will show that this paradox only occurs because we are using a second quantized theory to explain 
a third quantized phenomena. Thus, this information paradox does not occur 
 in third quantized loop quantum cosmology i.e., group field cosmology \cite{qw}-\cite{qw1}. 
In fact we will argue that an similar paradox would occur even if we used   first quantized quantum mechanics to 
analyse any second quantized phenomena. 

\section{First, Second and Third Quantization}
Before we analyse the occurrence of the black hole paradox, let us analyse a simple quantum mechanical system. 
Let us keep supplying energy to a particle.  
 Now as long as we keep the energy low enough,
 we can describe the dynamics of this particle  by using the single particle  quantum mechanics. 
  However, if we increase the energy to such a limit that new particles can be created, then we 
can not use single particle quantum mechanics.
This is because from the perspective of 
single particle quantum mechanics,  it would  appear that the wave function of this particle suddenly disappears, as soon as it decays 
into some other particles.  
Thus,  from the perspective of a single particle quantum mechanics, it would appear as an information paradox, as    
we will not be able to describe the system using its wave function.

This is even expected because a path of a single particle is 
topologically different from the path of two particles, and thus cannot be continuously deformed into it. 
So, we cannot continuously go from one particle to two particle by increasing energy. We rather need to start from a multi-particle 
quantum field theory, and then calculate the probability of a single particle spontaneously decaying into two particles. From the perspective 
of a single particle quantum mechanics it will appear as a non-unitary process. However, from the perspective of a second quantized field theory, 
it will be a perfectly unitary process. This is because the first quantized wave function would now be viewed as the classical field and the 
second quantized wave function would allow such particle decays. 
Thus, if we study second quantized phenomena in the framework  of the first quantized theories, we will get 
an apparent  information paradox.

A similar situation  occurs when we use a second quantized formalism to analyse 
a third quantized phenomena. The formation and evaporation of black hole involves changing the topology 
of spacetime and hence is a third quantized phenomena. If we study it using a   second quantized formalism, we 
will again get an apparent information paradox. 
In fact, in analogy with the above example of particles, a black hole is also topologically 
distinct from flat spacetime.  So, just like a single particle cannot be 
 continuously  deformed into multi-particles, a black hole cannot also  continuously evaporate  into flat spacetime. 
However,  an analysis of the decay of a black hole into flat spacetime can be done in a third quantized framework.
We will show that in  third quantized loop quantum gravity, there is no black hole information paradox, just like there is no 
apparent information paradox during particle annihilation in quantum field theory. 

Hawking radiation from a black hole is described by a second quantized quantum field theory, which is an approximation to third quantized gravity,
just like the single particle quantum mechanics is an approximation to  quantum field theory. This approximation holds till 
the black hole is big enough. However, as soon as the black hole approaches Planck size this approximation breaks down because 
 now topology can  change. 
At that stage we will have to use a third quantized formalism to explain the transition of a black hole into flat spacetime. 
 We will show that it is indeed possible to construct a unitarity third quantized 
theory of gravity accounting for topology change. 
In this theory, the decay of black hole will occur spontaneously to flat spacetime, in analogy with a spontaneous 
decays of particles into other particles in quantum field theory. 
 Thus, in reality there is no real information paradox, only the second quantized formalism used
to calculate the Hawking radiation breaks down at Planck scale. 

\section{Group Field Theory}
Now to  analyse the phenomena of information loss  properly, 
 we will have to use a third quantized formalism. We could analyse this phenomenon using 
third quantized canonical quantum gravity. However, as canonical quantum gravity has evolved into loop quantum gravity, and 
third quantization has evolved into group field cosmology; we will analyse this phenomenon using group field cosmology. 

In the loop quantum cosmology 
 the curvature of $A^i_\mu$ is expressed through the holonomy around a loop \cite{1vabdd}-\cite{1a}. The area of such a loop cannot 
cannot be smaller than a fixed minimum area because 
the smallest  eigenvalue of the area operator in loop quantum gravity is nonzero.  
Furthermore, the  eigenstates of the volume operator 
$\cal{V}$ are  
$ {\cal{V}} |\nu \rangle = 2 \pi \gamma G |\nu| |\nu \rangle
$, where $\nu = \pm a^2 {\cal{V}}_0 /2\pi \gamma G$ has the dimensions of length. 
So, in Planck units the Hamiltonian constrain for a homogeneous isotropic
 universe with a massless scalar field $\phi$,  can  be written as  \cite{a}-\cite{as}
\begin{equation} 
 K ^2\Phi(\nu, \phi) = [E^2 - \partial^2_\phi] \Phi(\nu, \phi) =0,
\end{equation}
where $\nu_0 =4$ and 
\begin{eqnarray}
 E^2 \Phi(\nu, \phi) &=& - [B(\nu)]^{-1}C^+(\nu) \Phi(\nu+4 , \phi)
 - [B(\nu)]^{-1}C^0(\nu)\Phi(\nu, \phi)  \nonumber \\ && - [B(\nu)]^{-1}C^-(\nu)\Phi(\nu-4, \phi).
\end{eqnarray}
Now, $K_1 = E$ and $K_2 = \partial_{\phi}$ and $\eta_{\mu\nu} =(1,-1)$, so we have $\eta^{\mu\nu}K_{\mu} K_{\nu}  = K^2$. 
Just as the wave function of first quantized theories is viewed as a classical field in second quantized formalism, the wave function of the 
second quantized theory is viewed as a classical field in third quantized formalism. Hence, the wave function of loop quantum cosmology 
will now be viewed as the classical field of group field cosmology. 
So, we can write the free part of the group field theory as \cite{qw}
\begin{equation}
 S_{free} = \sum_\nu \int d\phi \, \,    \Phi (\nu, \phi)  K^2  \Phi(\nu, \phi). 
\end{equation}
Now, we can also construct a complex field theory describing this model of loop quantum gravity. To do so we first construct 
covariant derivatives of the form, 
\begin{eqnarray}
{\cal{K}}_\mu \Phi (\nu, \phi) &=& K_\mu \Phi (\nu, \phi) + i {\cal{A}}_\mu (\nu, \phi)\Phi (\nu, \phi), \nonumber \\ 
{\cal{K}}_\mu \overline \Phi (\nu, \phi) &=& K_\mu\overline \Phi (\nu, \phi) 
- i {\cal{A}}_\mu (\nu, \phi)\overline\Phi (\nu, \phi),
\end{eqnarray}
and we define a field strength $ W_{\mu\nu} (\nu, \phi)$  as 
\begin{equation}
 W_{\mu\nu} (\nu, \phi) = K_\mu {\cal{A}}_\nu (\nu, \phi) - K_{\nu} {\cal{A}}_\mu (\nu, \phi).
\end{equation}
Now we can construct the action for the  complex group field theory  as follows, 
\begin{eqnarray}
  S_{c}&=& \sum_\nu \int d\phi \, \,   \overline \Phi (\nu, \phi) {\cal{K}}^2  \Phi(\nu, \phi) \nonumber \\ && 
-  \sum_\nu \int d\phi \, \,  W^{\mu\nu}(\nu, \phi) W_{\mu\nu} (\nu, \phi). 
\end{eqnarray}
This action is invariant under the following gauge transformations, 
\begin{eqnarray}
 \Phi (\nu, \phi)= i\Lambda ( \nu, \phi) \Phi (\nu, \phi), &&
 \overline\Phi (\nu, \phi)=-i \overline\Phi (\nu, \phi)\Lambda ( \nu, \phi) , \nonumber \\
{\cal{A}}_\mu(\nu, \phi)=  K_\mu \Lambda ( \nu, \phi),&&
\end{eqnarray}
where $\Lambda ( \nu, \phi)$ is a group scalar  field in its own right. 
\section{Quantization}
As the classical  action for the group field cosmology has a gauge symmetry, we cannot quantize it without fixing a gauge. 
We thus fix the following gauge, 
\begin{equation}
 K^\mu{\cal{A}}_\mu (\nu, \phi) =0, 
\end{equation}
To impose this condition at a quantum level, we add the following gauge fixing  and ghost terms \cite{qw2},
\begin{eqnarray}
 S_{gh} &=& \sum_\nu \int d\phi \, \,    s \, B  ( \nu, \phi )  K^\mu{\cal{A}}_\mu ( \nu, \phi ), \nonumber \\ 
 S_{gf} &=& \sum_\nu \int d\phi \, \,    s \, \overline{C}  ( \nu, \phi )  K^\mu {\cal{K}}_\mu C ( \nu, \phi ),
\end{eqnarray}
to the original classical action. The sum of the gauge fixing term and the ghost term can be expressed as 
\begin{equation}
  S_{gh} + S_{gf} = \sum_\nu \int d\phi \, \,    s \, \overline{C}  ( \nu, \phi )  K^\mu{\cal{A}}_\mu ( \nu, \phi ) ,
\end{equation}
where the third quantized BRST transformations are given by 
\begin{eqnarray}
  s\,\Phi(\nu, \phi )=i C (\nu, \phi )\Phi(\nu, \phi ) , && s\, B( \nu, \phi ) =0,\nonumber \\ 
s\, \overline  \Phi (\nu, \phi ) = -i \overline \Phi(\nu, \phi ) C(\nu, \phi ),  &&
s\, C( \nu, \phi )=0, \nonumber \\
 s\, {\cal{A}}_\mu ( \nu, \phi ) =  K_\mu  C( \nu, \phi ), 
&&
s\, \overline{C} ( \nu, \phi ) = B( \nu, \phi ).
\end{eqnarray}
 These third quantized BRST 
transformations are nilpotent, $s^2 =0$, and so the sum of the ghost term with the gauge fixing term is invariant under these third quantized 
BRST transformations.
As the third quantized BRST transformations of the original third quantized classical action is only a ghost valued gauge transformations, 
it is also invariant
under these third quantized BRST transformations. Thus, the full action is invariant under these transformations, $
 s\, S  =0$, where $S =  S_{c}  + S_{gf} + S_{gh}$.
 These third quantized BRST transformation can be used to construct a third quantized BRST charge and analyse
 the unitarity of the  group field 
cosmology with gauge symmetry. 

To construct the Fock space, we expand these fields into modes which we promote to operators. 
Each of these will create or annihilate geometry and matter. 
The vacuum state $|0\rangle$ is defined by, 
$
 a_k |0\rangle =0,
$
and the Fock space is constructed  by the action of creation operators on this vacuum state. 
The vacuum state here corresponds to a state with no
geometry, matter field and topological structure. The topology, geometry and matter is created by the action of 
creation operators on this vacuum state. Interactions   correspond to an interaction of these universes. 
  Thus, the topology changing 
processes will now occur as the action contains  interaction terms.

As the total Lagrangian  density   
is  invariant under the  BRST   transformations, so 
 we can obtain  
the Norther's charge  
corresponding to the BRST transformations and use 
it to project out the physical state of the theory. 
Now as the matter fields act as the time variable, we define our Norther's  charge as 
\begin{eqnarray}
Q( \phi)  & = &   \sum_\nu \,\, 
\left[ \frac{ \partial L_{eff}  }{\partial \partial_{\phi} {\cal{A}}_\mu (\nu, \phi )   } 
 s\, 
{\cal{A}}_\mu (\nu, \phi )   +
 \frac{ \partial L_{eff}  }{\partial  \partial_{\phi} C (\nu, \phi )   }  s\, C (\nu, \phi )  
 \right.\nonumber \\ &&   +
\frac{ \partial L_{eff}  }{\partial  \partial_{\phi} \overline{C} (\nu, \phi )   }
 s\, \overline{C}  (\nu, \phi )  
 +
\frac{ \partial L_{eff}  }{\partial \partial_{\phi} B  (\nu, \phi )  }  s\, B (\nu, \phi ) 
\nonumber \\ && \left.  + 
\frac{ \partial L_{eff}  }{\partial \partial_{\phi} \Phi  (\nu, \phi )   }  s\, 
 \Phi (\nu, \phi )    +
 \frac{ \partial L_{eff}  }{\partial \partial_{\phi}\overline \Phi  (\nu, \phi )   } 
 s\, 
\overline \Phi  (\nu, \phi ) \right], 
\end{eqnarray}
where 
\begin{equation}
  S = \sum_{\nu} \int D\phi \,\, L_{eff}. 
\end{equation}
As the BRST transformations are nilpotent, 
so we have, $
 Q^2  = 0
$.
The  physical states  $ |phy \rangle $ can now be defined as  
states that are annihilated by $Q$, 
$
 Q |phy \rangle =0. 
$
The asymptotic states are given by 
\begin{eqnarray}
 |{out}\rangle &=& |\nu_1, \phi_1\rangle, \nonumber \\
 |{in}\rangle &=& |\nu_2, \phi_2\rangle,
\end{eqnarray}
where $ |{pa,out}\rangle$ is the state before the formation of the black hole and $ |{pb,in}\rangle$ is the state after the evaporation  of the 
black hole. 
A  $\mathcal{S}$-matrix element between them can be written as
\begin{equation}
\langle{out}|{in}\rangle = \langle{pa}|\mathcal{S}^{\dagger}\mathcal{S}|{pb}\rangle.
\end{equation}
As the      BRST   charge is   
a conserved charge, so  the evolution of any physical state will 
also be annihilated by it, 
$
 Q \mathcal{S} |phy\rangle =0.
$ 
This implies that the states $\mathcal{S}|pb\rangle$ must be a linear combination of physical states $|a_{0, i}\rangle$, 
\begin{equation}
\langle pa|\mathcal{S}^{\dagger} \mathcal{S}|pb \rangle
 = \sum_{i}\langle pa|\mathcal{S}^{\dagger}|a_{0, i}\rangle
\langle a_{0, i}| \mathcal{S}| pb \rangle.
\end{equation}
Since the full $\mathcal{S}$-matrix is unitary this relation implies that the  $S$-matrix restricted to
physical sub-space is also unitarity. Hence, the  topology changing process are unitarity. 
So, in third quantized loop quantum gravity, there is no information loss. 
\section{Conclusion}
In this paper we argued that the information paradox only occurs because we are using a second quantized formalism for analysing a 
third quantized phenomena. A similar paradox would also occur if we used single particle quantum mechanics for 
analyzing a multi-particle system at energies at which new particles would be created. 
Thus, in reality there is no information paradox but only a breakdown of second quantized formalism. In this paper
we first constructed a
gauged version of group field cosmology and then studied its BRST symmetry. 
The BRST charge thus calculated was used for projecting out the physical states of the theory.
It was thus demonstrated that the $S$-matrix is unitary. 
This theory naturally accounted for process involving topology change. 
Hence, it was argued that such processes are unitary in third quantized loop quantum gravity. In particular, 
it was shown that the creation  and annihilation  of a black hole is a unitary 
process in group field cosmology. It will be interesting to analyse the big bang in this formalism. 
Just like the annihilation of  a black hole is explained  in group field cosmology, similarly the creation 
of universe can also be explained in it. It will be interesting to perform this analysis and explicitly show 
how the big bang is naturally accounted for in group field cosmology.

\end{document}